\documentstyle{article}
\begin{document}
\title{An exterior for the G\"{o}del spacetime}
\author{W.B.Bonnor, Queen Mary and Westfield College, London E1 4NS,\\
N.O.Santos, Observat\'{o}rio Nacional, 20921, Rio de Janeiro, Brazil,\\
M.A.H.MacCallum, Queen Mary and Westfield College, London E1 4NS}
\maketitle
\setlength{\parindent}{0.5in}
\begin{abstract}
We match the vacuum, stationary, cylindrically symmetric solution of
Einstein's field equations with $\Lambda$,
in a form recently given by Santos, as
an exterior to an infinite cylinder  of dust cut out of a G\"{o}del
universe.  There are three cases, depending on the radius of the cylinder.
Closed timelike curves are present in the exteriors of some of the solutions.
There is a considerable similarity between the spacetimes investigated here
and those of van Stockum referring to an infinite cylinder of rotating
dust matched to vacuum, with $\Lambda=0$.
\end{abstract}
\section{Introduction}
van Stockum [1] solved the problem of a rigidly rotating infinitely long cylinder
of dust, using Einstein's equations without cosmological constant.  The metric for the
interior is unique, depending on one parameter $a$; but the vacuum exterior
has three cases, depending on the mass per unit length of the interior.
The metric corresponding to the case of lowest mass can be diagonalised
locally (but not globally), but this is not possible for the other
two cases.  (See [2] and references given therein.)

In this paper we consider a similar problem using Einstein's equations
with negative cosmological constant.  For the interior containing rotating matter
we use an infinite cylinder cut out of a G\"{o}del universe.  The exterior
metric has been given by various authors, notably Krasi\'{n}ski [4]
(see [9] for more information).  However, in this work we use it in the form
recently obtained by Santos [3]; it satisfies
\begin{equation}
R^{ik}-\frac{1}{2}g^{ik}R=\Lambda g^{ik},
\end{equation}
It turns out that in this problem too the
exterior has three cases, depending on the radius (and therefore mass per unit length) of the cylinder.

\section{The metrics}
For the interior we use the G\"{o}del metric in the form
\begin{equation}
ds^{2}=dR^{2}+dZ^{2}+4b^{2}(\sinh^{2}\rho-\sinh^{4}\rho)d\psi^{2}
-4(2^{\frac{1}{2}})b\sinh^{2}\rho\, d\psi\,dT -dT^{2},
\end{equation}
where
\[\rho=R/2b,\;\;\;b=(-2\Lambda)^{-1/2}.\]
{\em Throughout the paper positive square roots are to be taken unless the contrary
is indicated}.
The ranges of the coordinates are
\[R \leq R_{0},\;-\infty<Z<+\infty,\;0 \leq \psi \leq 2\pi,\;-\infty<T<+\infty,\]
and $\psi=0$ and $\psi=2\pi$ are to be identified.

The exterior metric is that of [3] for $\Lambda<0$ in slightly different form:
\begin{equation}
ds^{2}=dr^{2}+\exp{\mu}\,dz^{2} +ld\phi^{2} +2k\,d\phi\,dt-fdt^{2},
\end{equation}
$\mu,l,k,f$ being functions of $r$ only.
We shall write
\begin{equation}
D^{2}=lf+k^{2},\;\;G=e^{\mu/2}D,
\end{equation}
and suppose that $D$ and $G$ are positive.
Then the complete solution can be written
\begin{eqnarray}
G&=&C_{1}\cosh \sqrt{3/2}(r/b)+C_{2}\sinh\sqrt{3/2}(r/b),\\
e^{3\mu/2}&=&\epsilon G\exp(\delta \Theta/\gamma),\\
l&=&-D\alpha^{-1} \sinh(\alpha \Theta/2\beta),\\
k&=&D[\cosh(\alpha \Theta/2\beta)+(\beta/\alpha)\sinh(\alpha \Theta/2\beta)],\\
f&=&D[2\beta \cosh(\alpha \Theta/2\beta)+\alpha^{-1}(\alpha^{2}+\beta^{2})
   \sinh(\alpha \Theta/2\beta)],\\
\Theta^{\prime}&=&\gamma /G,
\end{eqnarray}
where $\alpha, \beta, \gamma, \delta, \epsilon, C_{1}, C_{2}$ are constants,
all real except $\alpha$ which can be real or imaginary; they must
satisfy one relation
\begin{equation}
\alpha^{2}\gamma^{2}/ \beta^{2}=8b^{-2}(C_{2}^{2}-C_{1}^{2}) -4\delta^{2}/3;
\end{equation}
a prime, as in (10), means $d/dr$.
Another constant, not occurring in (11), arises in the integration of (10).
One can easily check that (7),(8),(9) give real, finite expressions if $\alpha$ is
imaginary or is allowed to tend to zero.

The issue of the constants arising in this solution is discussed in another paper
and here we shall simply show how they can be fixed by matching the
solution to the G\"{o}del metric.  The same matching procedure applies
whether we take the G\"{o}del metric as interior or exterior.  We choose the former
because it yields a globally regular solution, whereas with a G\"{o}del
exterior and Santos interior it seems that we must allow a singularity
along the symmetry axis.

\section{Matching the metrics}
We wish to match the metrics (2) and (3) across a hypersurface $\Sigma$
which is $R=R_{0}$ in (2) and $r=r_{0}$ in (3) ($R_{0},r_{0}>0)$.
With respect to this
hypersurface both the metrics are in Gaussian form, and we assume that on it
the coordinates $(Z,\psi,T)$ and $(z,\phi,t)$ are the same and have
the same ranges.  Then the Lichnerowicz matching conditions will be satisfied
if we equate on $\Sigma$ the metric components and also their first derivatives.
This means that $\mu, k,l,f$ and their derivatives need to take the
corresponding values from the metric (2).  For the time being we exclude
the cases in which
\begin{equation}
\sinh^{2}\frac{R_{0}}{2b}=\frac{1}{2};\;\;\sinh^{2}\frac{R_{0}}{2b}=1
\end{equation}

We shall consider the matching problem in the form
\begin{eqnarray}
                         &e^{\mu}                                                               & 1\\
                         &\mu^{\prime}                                                               & 0\\
k^{2} +fl               =&G^{2}e^{-\mu}                                                               & b^{2}\sinh^{2}2\rho_{0} \\
f^{\prime}l-l^{\prime}f =&G^{2}e^{-\mu}\Theta^{\prime}                                   & -2b\sinh2\rho_{0}(1-2\sinh^{2}\rho_{0})\\
f^{\prime}k-k^{\prime}f =&G^{2}e^{-\mu}\Theta^{\prime}(\alpha^{2}-\beta^{2})/2\beta      & 2^{\frac{1}{2}}\sinh2\rho_{0}\\
k^{\prime}l-l^{\prime}k =&G^{2}e^{-\mu}\Theta^{\prime}/2\beta                            & -4(2^{\frac{1}{2}})b^{2} \sinh2\rho_{0}\sinh^{4}\rho_{0}\\
k/l                     =&-\alpha\coth(\alpha\Theta/2\beta)-\beta                        & -[2^{\frac{1}{2}}b(1-\sinh^{2}\rho_{0})]^{-1}\\
f/l                     =&-2\alpha\beta\coth(\alpha\Theta/2\beta)-(\alpha^{2}+\beta^{2}) & [4b^{2}(\sinh^{2}\rho_{0}-\sinh^{4}\rho_{0})]^{-1}\\
                         &f                                                              & 1
\end{eqnarray}
where $\rho_{0}:=R_{0}/2b$.  The quantities in the right-hand column are
obtained from (2), and give the values which must be assumed by the quantities
in the left-hand column.  It is not hard to show that this set of equalities
solves the matching problem as stated in the previous paragraph.  Eqn (21) is
necessary because (15)-(20) do not determine the sign of the set $[k,l,f,
k^{\prime},l^{\prime},f^{\prime}]$, so the sign of one member of the set must be prescribed.
In this form, which has been chosen for ease of calculation, there is a
redundancy: nine equations for eight conditions arising from the continuity
of $\mu,k,l,f$ and their derivatives.

By a shift of the origin of $r$ (which amounts to a redefinition of $C_{1}, C_{2}$ in
(5)) we can arrange that the radial coordinate is
continuous on $\Sigma$, so $r_{0}=R_{0}$.  From the continuity of $D$
and its derivative, and $\mu$ and its derivative, we deduce the continuity of
$G$ and its derivative.  On $\Sigma$ we have
\begin{equation}
G=b\sinh 2\rho_{0},\;\;G^{\prime}=\cosh 2\rho_{0},
\end{equation}
so, using (5) we find
\begin{eqnarray*}
C_{1}\cosh\sqrt{6}\rho_{0}+C_{2}\sinh\sqrt{6}\rho_{0}&=&b\sinh2\rho_{0},\\
C_{1}\sinh\sqrt{6}\rho_{0}+C_{2}\cosh\sqrt{6}\rho_{0}&=&\sqrt{\frac{2}{3}}b\cosh2\rho_{0},\\
\end{eqnarray*}
whence
\begin{eqnarray}
C_{1}=b[\sinh2\rho_{0}\cosh\sqrt{6}\rho_{0}-\sqrt{\frac{2}{3}}\cosh2\rho_{0}\sinh\sqrt{6}\rho_{0}],\\
C_{2}=b[\sqrt{\frac{2}{3}}\cosh2\rho_{0}\cosh\sqrt{6}\rho_{0}-\sinh2\rho_{0}\sinh\sqrt{6}\rho_{0}].
\end{eqnarray}

We can now write down an expression for $G$ in terms of $r,r_{0}$ and $b$.
From (5), (23) and (24) we find, after a short calculation
\begin{equation}
G=b[\sinh2\rho_{0} \cosh \sqrt{\frac{3}{2}}\frac{r-r_{0}}{b} +\sqrt{\frac{2}{3}}
\cosh2\rho_{0}\sinh\sqrt{\frac{3}{2}}\frac{r-r_{0}}{b}],
\end{equation}
which shows that $G$ is positive for $r\geq r_{0}$ as required by (4).

We now proceed with matching (13)-(20), excluding for the time being the two
cases (12).
The arbitrary constant $\epsilon$
can be chosen to satisfy (13), but the continuity of $\mu^{\prime}$ in (14)
requires
\begin{equation}
G^{\prime}_{0} +\delta =0,
\end{equation}
where the suffix $0$ means the value on $\Sigma$.  Using (22) we find
\begin{equation}
\delta=-\cosh 2\rho_{0}.
\end{equation}
(15) is satisfied in virtue of (13) and (22).  (16)-(18) lead to
\begin{eqnarray}
(\Theta^{\prime})_{0}&=&-(b\cosh\rho_{0}\sinh\rho_{0})^{-1}(1-2\sinh^{2}\rho_{0}),\\
\beta              &=&(4\sqrt{2}b\sinh^{4}\rho_{0})^{-1}(1-2\sinh^{2}\rho_{0}),\\
\alpha^{2}         &=&(32b^{2}\sinh^{8}\rho_{0})^{-1}Y^{2},
\end{eqnarray}
where $Y^{2}=1-4\sinh^{2}\rho_{0}-4\sinh^{4}\rho_{0}$.
The case in which $\alpha$ is  zero will be considered in Section 4.  If
$\alpha\neq 0$
(19) now gives
\begin{equation}
\alpha\coth\frac{\alpha\Theta_{0}}{2\beta}=-\frac{1-3\sinh^{2}\rho_{0}-2\sinh^{4}\rho_{0}}
{4\sqrt{2}b\sinh^{4}\rho_{0}(1-\sinh^{2}\rho_{0})},
\end{equation}
and from (29), (30) and (31) we find that (20) is satisfied.

To verify (21) we first calculate from (31)
\begin{equation}
\alpha^{-1}\sinh\frac{\alpha\Theta_{0}}{2\beta}=2b\eta\tanh\rho_{0}(1-\sinh^{2}\rho_{0}),
\end{equation}
where $\eta=\pm 1$.  Using (9) we now write $f$ in the form
\begin{equation}
f=D\alpha^{-1}\sinh\frac{\alpha\Theta}{2\beta}[2\alpha\beta\coth
\frac{\alpha\Theta}{2\beta}+\alpha^{2}+\beta^{2}],
\end{equation}
so that, on the boundary we have after a calculation using (15),(28),(29),(31) and (32)
\begin{equation}
f_{0}=-\eta.
\end{equation}
Therefore to satisfy (21) we must take
\begin{equation}
\eta=-1.
\end{equation}
Because of (32) and (34) $\Theta_{0}$ must satisfy
\begin{equation}
\sinh\frac{\alpha\Theta_{0}}{2\beta}=-2b\alpha \tanh\rho_{0}(1-\sinh^{2}\rho_{0}),
\end{equation}
where from (30)
\begin{equation}
\alpha=\pm(4\sqrt{2}b\sinh^{4}\rho_{0})^{-1}(1-4\sinh^{2}\rho_{0}-4\sinh^{4}\rho_{0})^{1/2},
\end{equation}
both signs on the right being allowed.

Finally we must check that the constants as determined above satisfy (11).
First from (10), (22) and (28) we get
\begin{equation}
\gamma=-2(1-2\sinh^{2}\rho_{0}),
\end{equation}
and inserting this, together with the other constants into (11) we find it
satisfied.

{\em We have matched Santos's metric (3) at $r=R_{0}$ as an exterior to the G\"{o}del metric
except for the special cases in which $\sinh^{2}\rho_{0}$ is equal to $1$ or $1/2$.}
It will now be convenient to consider separately the three cases determined
by the sign of $\alpha^{2}$.

\section{The various cases}
We now explore the various cases that can arise, paying special attention to
the occurrence of closed timelike curves (CTC).  We make use of a theorem of
Carter [6,7] (cf. Tipler [8]). It states that in any spacetime with an Abelian
iso\-metry group which is everywhere transitive on timelike surfaces, any
point can be connected to any other point by both a future and a past
timelike curve (and hence a CTC) if and only if there is no Lie algebra
covector whose corresponding differential form is everywhere well-behaved
and timelike (Carter refers to this form as spacelike or null, but has in
mind the nature of its orthogonal hypersurface: we shall not discuss the
null case). For the metric (3), the Lie algebra covectors are real-valued
linear maps, with constant coefficients, on the space of vectors of the
form $v^t\partial_{t} + v^\phi \partial_{\phi} + v^z \partial_{z}$,
so we have to examine the nature of a one-form ${\bf w}=Adt +Bd \phi +Cdz$,
with constant $A$, $B$ and $C$.  Such a one-form has $g^{ab}w_{a}w_{b}
=(-lA^{2}+2kAB+fB^{2})/D^{2} +C^{2}\exp {(-\mu)}$, and we want to find out
whether or not ${\bf w}$ can be timelike everywhere, i.e. satisfy
$g^{ab}w_{a}w_{b}<0$ for all $r$.  Since the contribution from $C$
is always positive, we need consider only the case $C=0$.  If $l>0$ for all
$r$, ${\bf w}=dt$ is clearly timelike, so there exist no CTC.  On the
other hand, if $l<0$ for some $r$, the circles on which $r$, $z$ and $t$
are constant are clearly CTC.  The outcome is that in our spacetimes,
excluding the possibility that $l \geq 0$ but $l=0$ for some $r$, there
are CTC if and only if $l<0$ for some $r$.\vspace{.2in}
\newline
{\bf 4.1 The case $\alpha^{2}>0$}\\
The condition for this is $Y^{2}>0$, i.e.
\begin{equation}
\sinh^{2}\rho_{0}<\frac{1}{2}(\sqrt{2}-1) \approx 0.207,
\end{equation}
and if it is fulfilled $\sinh\frac{\alpha\Theta}{2\beta}$ is real and
the functions in the solution are hyperbolic with real argument.

We consider the exterior in this case to see whether it contains
closed time-like curves, i.e. whether  $l<0$ in $r>r_{0}$.
Because $l$ is continuous on $\Sigma$ we know that it is positive on $r=r_{0}$;
we shall show that it remains positive for $r>r_{0}$.  We first note from (10),(25) and
(38) that $\Theta^{\prime}<0$ for $r>r_{0}$.  From this it follows that
\begin{equation}
\frac{d}{dr}(\alpha^{-1}\sinh\frac{\alpha\Theta}{2\beta})=
\frac{\Theta^{\prime}}{2\beta}\cosh\frac{\alpha\Theta}{2\beta},
\end{equation}
is negative for $r>r_{0}$, because from (29) $\beta>0$.   Hence $\alpha^{-1}\sinh(\frac{\alpha\Theta}{2\beta})$
diminishes from the negative value it has when $\rho=\rho_{0}$, given by (36).
Thus from (7) $l>0$ in $r>r_{0}$, and CTC do not exist.

There are no CTC in the G\"{o}del interior either because in (2) $l>0$
for $R<R_{0}$ in this case.\vspace{0.2in}\\
{\bf 4.2 The case $\alpha=0$}\\
In this case
\begin{equation}
\sinh^{2}\rho_{0}=\frac{1}{2}(\sqrt{2}-1)\approx 0.207.
\end{equation}
This value of $\rho_{0}$ is critical in that it defines the only radius at
which the G\"{o}del model admits circular null geodesics.
The functions $f,k,l$ of (3) are given by
\begin{eqnarray*}
l&=&-D\Theta/2\beta,\\
k&=&D(2+\Theta)/2,\\
f&=&D(4\beta+\beta\Theta)/2.
\end{eqnarray*}
$\beta$ is still given by (29) which yields, when (41) is used,
\[\beta = b^{-1}(1+\sqrt{2}).\]
It is easy to show that the value of $\Theta$ on $\Sigma$ is negative, namely
\[\Theta_{0}=-2(3-\sqrt{2}),\]
and also that
\[\gamma=-2(2-\sqrt{2}),\]
so from (10) $\Theta^{\prime}<0$ for $r>r_{0}$.  Now an argument similar to that
used in the case $\alpha^{2}>0$ shows that $l>0$ in $r>r_{0}$ so there
are no CTC in the exterior.  There are no CTC in the interior either.\vspace{0.2in}\\
{\bf 4.3 The case $\alpha^{2}=-a^{2}<0$}\\
The hyperbolic functions in (7), (8) and (9) become trigonometric:
\begin{eqnarray}
l&=&-Da^{-1}\sin(a\Theta/2\beta),\\
k&=&D[\cos(a\Theta/2\beta)+(\beta/a)\sin(a\Theta/2\beta)],\\
f&=&D[2\beta\cos(a\Theta/2\beta)+a^{-1}(\beta^{2}-a^{2})\sin(a\Theta/2\beta)],
\end{eqnarray}
where $a$ is real, and the constants have to satisfy
\begin{equation}
8b^{-2}(C_{2}^{2}-C_{1}^{2})=4\delta^{2}/3-a^{2}\gamma^{2}/\beta^{2}.
\end{equation}

To investigate the occurrence of CTC when $\alpha^{2}<0$ we must first integrate (10),
$G$ being given by (25).  This leads to considerable mathematical complexity,
so we shall restrict ourselves to four special cases. \vspace{0.1in}
\newline
{\bf 4.3.1 The sub-case $\tanh2\rho_{0}=\sqrt{2/3}\leftrightarrow \sinh^{2}\rho_{0}=\frac{1}{2}(\sqrt{3}-1)\approx 0.366.$}\\
For this value of $\rho_{0}$ we have $\sinh^{2}\rho_{0}=\frac{1}{2}(\sqrt{3}-1), Y=\pm i$, and from (29), (37) and (38)
\begin{equation}
\alpha=\pm i(2\sqrt{2}b)^{-1}(2+\sqrt{3}),\;\beta=(2\sqrt{2}b)^{-1},\;\gamma=-2(2-\sqrt{3}).
\end{equation}
(25) gives
\begin{equation}
G=\sqrt{2}b\exp{\sqrt{\frac{3}{2}}\frac{r-r_{0}}{b}},
\end{equation}
and (10) yields
\begin{equation}
\Theta=\frac{2}{3}(2\sqrt{3}-3)\exp{[-\sqrt{\frac{3}{2}}\frac{r-r_{0}}{b}]} +C,
\end{equation}
where $C$ is a constant of integration.  On $r=r_{0}$ we have
\begin{equation}
\Theta=\Theta_{0}=\frac{2}{3}(2\sqrt{3}-3)+C,
\end{equation}
and at $r=\infty$
\begin{equation}
\Theta_{\infty}=C.
\end{equation}
Our aim is to find whether $l$, given by (42), which is positive on $r=r_{0}$, becomes negative as $\Theta$
ranges between these two values: if so, this will denote the existence of CTC.

We now insert the above values of $\sinh\rho_{0},Y,\alpha,\beta$ into (36)
and find after a short calculation
\begin{equation}
\sin[(1+\frac{1}{2}\sqrt{3})\Theta_{0}]=-\frac{\sqrt{3}}{2},\;\;
\cos[(1+\frac{1}{2}\sqrt{3})\Theta_{0}]=-\frac{1}{2},
\end{equation}
whence
\begin{equation}
(1+\frac{1}{2}\sqrt{3})\Theta_{0}=\frac{4\pi}{3}+2n\pi,
\end{equation}
where $n$ is a positive or negative integer or zero.  We now substitute $\alpha$ and $\beta$
from (46) and $\Theta_{\infty}$ from (50) into (42) to get the value of $l$ at $r=\infty$:
\begin{equation}
l_{\infty}=-\frac{2\sqrt{2}bD}{2+\sqrt{3}}\sin(1+\frac{1}{2}\sqrt{3})C,
\end{equation}
so, using (49) and (52) we have
\begin{equation}
l_{\infty}=-\frac{2\sqrt{2}bD}{2+\sqrt{3}}\sin[(\frac{4\pi}{3}+2n\pi)-\frac{\sqrt{3}}{3}].
\end{equation}
For any $n$ this sine is positive, so $l_{\infty}$ is negative and CTC exist
in the exterior of the cylinder.  They do not exist in the interior for this radius. \vspace{0.1in}
\newline
{\bf 4.3.2 The sub-case $\sinh^{2}\rho_{0}=\frac{1}{2}$}\\
Although $\alpha^{2}$ given by (37) is negative in this case, the solution
(42)-(44) does not apply because
with this special value of $\sinh^{2}\rho_{0}$ it follows from (16) that
$\Theta ^{\prime}=0$ on $\Sigma$ so from (10) $\gamma=0$ and $\Theta$ is
constant throughout the exterior.  This leads to a case not included in [3],
namely, that in which $f$ and $l$ are proportional.  The constant of proportionality
is found from (20) to be $b^{-2}$ so we have
\begin{equation}
f=b^{-2}l.
\end{equation}

Let us introduce a function $\Phi$ as in (15), (17) of [3] by
\begin{equation}
\Phi^{\prime}=v^{\prime}(b^{-2}+v^{2})^{-1},\;\;v=k/l;
\end{equation}
then we find
\begin{equation}
f=b^{-1}D\cos b^{-1}\Phi,\;k=D\sin b^{-1}\Phi,\;l=bD\cos b^{-1}\Phi,
\end{equation}
where
\begin{equation}
\Phi^{\prime}=\nu G^{-1},
\end{equation}
$\nu$ being a constant.\footnote{(58) follows from (18) of [3], but because of
the different $r$ coordinate implied by (1) of [3] and (4) of this paper we
must write $G$ instead of $D$.}

The constant $\nu$ can be determined from (58) by writing $\Phi^{\prime}$ as
\begin{equation}
\Phi^{\prime}=\frac{k^{\prime}l-l^{\prime}k}{D^{2}}
\end{equation}
and substituting on the right hand side from (18), (4) and (22); thus we find
\begin{equation}
\nu =-\sqrt{2}b.
\end{equation}

The next step is to integrate (58), inserting the value of $G$ obtained from (25)
after putting $\sinh \rho_{0}=\sqrt{1/2}$, namely
\[G=\frac{b}{\sqrt{3}}\cosh(\sqrt{\frac{3}{2}}\frac{r-r_{0}}{b} +\zeta)\]
where $\sinh\zeta=2\sqrt{2}$. Inserting this and (56) into (54) and integrating
we get
\begin{equation}
\Phi=-4b\tan^{-1}[\tanh \frac{1}{2}(\sqrt{\frac{3}{2}}\frac{r-r_{0}}{b}+\zeta)]+E,
\end{equation}
where E is a constant of integration.

Eqns (57) and (58) are what (7)-(10) become in this special case; of the remaining equations in the complete solution,
(5) is unchanged, (6) becomes
\begin{equation}
e^{3\mu/2}=\epsilon G e^{\Phi\delta/\nu},
\end{equation}
and in place of (11) we have
\begin{equation}
2(C_{2}^{2}-C_{1}^{2})-\frac{1}{3}\delta^{2}b^{2}=-\nu^{2}.
\end{equation}
One can check that this solution matches the G\"{o}del metric on $r=r_{0}$
provided $\delta=-2$, and $E$ is given an appropriate value as described
below.

We turn to the question whether there are CTC in the exterior in this case.
Equating $f,k,l$ in (57) to their values on $\Sigma$ as given by (2) with
$\rho=\rho_{0}$, and using $\sinh^{2}\rho_{0}=1/2$, we find
\begin{equation}
\Phi_{0}=b(-\varpi+2n\pi),
\end{equation}
where $n$ is an integer and $\varpi$ denotes the principal value of $\cos^{-1}(3)^{-1/2}$,
which is 0.955 radians, to three decimal places.  From (61)
\begin{equation}
\Phi_{0}=-4b \tan^{-1}(\tanh \frac{\zeta}{2}) + E;
\end{equation}
using $\sinh \zeta=2\sqrt{2}$ and (64) we get
\begin{equation}
E=\Phi_{0}+4b\tan^{-1}(2)^{-1/2}=b[-\varpi+4\theta +(4m+2n)\pi],
\end{equation}
$m$ denoting an integer and $\theta$ the principal value of $\tan^{-1}(2)^{-1/2}$,
which is 0.615.
On the other hand, putting $r=\infty$ in (61) we obtain
\begin{equation}
\Phi_{\infty}=-4b\tan^{-1}(1) + E =E-b(4s+1)\pi,
\end{equation}
where $s$ is an integer.
Eliminating $E$ we finally obtain
\begin{equation}
b^{-1}\Phi_{\infty}=(4\theta-\pi-\varpi) +2\pi(n+2m-2s).
\end{equation}

$l$ will change sign in $r>r_{0}$ if $\cos b^{-1}\Phi$ does so.  On $\Sigma,$
$b^{-1}\Phi=b^{-1}\Phi_{0}$ is in the fourth quadrant, as is clear from (64).
However, from (68) $b^{-1}\Phi_{\infty}$ is in the third quadrant, so
$\cos b^{-1}\Phi$ changes sign between $r=r_{0}$ and $r=r_{\infty}$
and so therefore does $l$.  Hence CTC must exist in the exterior spacetime,
though, as in {\bf 4.3.1}, they do not exist in the interior. \vspace {0.1 in}
\newline
{\bf 4.3.3 The sub-case $\sinh^{2}\rho_{0}=1$}\\
This is the second exceptional case in (12); it has
$l=0$ on $\Sigma$, so the $\phi$-coordinate circles are null there.  The exterior solution (5)-(11) still applies,
and so do the boundary conditions (13)-(21) except for (19) and (20).
From (4), (13) and (15) we have $D_{0}=2\sqrt{2}b$ and the continuity
of $g_{\phi\phi}$ and $g_{\phi t}$ on $\Sigma$ require
\begin{equation}
\sinh(\alpha \Theta_{0}/2\beta)=0,\;\cosh(\alpha \Theta_{0}/2\beta)=-1,
\end{equation}
whence
\begin{equation}
\frac{\alpha \Theta_{0}}{2\beta}=(2n+1)i\pi,
\end{equation}
where $n$ is an integer.
From the requirement that $g_{tt}=1$ on $\Sigma$ we find,
using (9) and (70)
\begin{equation}
\beta=-(4\sqrt{2}b)^{-1}.
\end{equation}
The other constants of the solution are now easily determined from (14)-(18) and (27):
\begin{equation}
\alpha^{2}=-7/(32b^{2}),\;(\Theta^{\prime})_{0}=(\sqrt{2}b)^{-1},\;\delta=-3,
\end{equation}
and all boundary conditions are satisfied on $\Sigma$.

That CTC exist in the exterior can be seen as follows.  As $D_{0}>0\,D$ will
be positive in the neighbourhood of $\Sigma$, from continuity.  Moreover, from (69), (71)
and (72) $\alpha^{-1}\sinh(\alpha\Theta_{0}/2\beta)$ is positive in the exterior
neighbourhood of $\Sigma$; hence using (42) we see that $l$ decreases from its
zero value on $\Sigma$ and CTC must exist in the \vspace{0.1in} exterior.\\
{\bf 4.3.4 The sub-case $\sinh\rho_{0}>1$}\\
In this case $l$ is negative within the cylinder, on its boundary, and, by continuity,
in the exterior at least near the boundary.  Therefore CTC exist inside and
outside the cylinder.

\section{Conclusion}
We have been studying a spacetime satisfying Einstein's equations with
negative cosmological constant, describing an infinite cylinder cut
from the G\"{o}del universe, surrounded by empty space.  The spacetime depends
on two parameters, $R_{0}(=r_{0})$, the coordinate radius of the cylinder,
and the cosmological constant.
Our solution of the matching
problem on the boundary of the cylinder shows that the exterior metric is
of three different types, depending on the radius of the cylinder.
For smaller radii there are no closed timelike curves (CTC) inside the cylinder
or in the empty exterior.  We have not given a complete treatment of CTC
for cylinders with larger radii, but we have shown that for some radii
they exist outside the cylinder even though there is none inside.  For sufficiently large
radii ($\sinh \rho_{0}>1$) CTC exist in the interior and the exterior.

There are similarities between our results and the corresponding ones for the van Stockum spacetimes,
which refer to infinite dust cylinders in vacuum with $\Lambda=0$.  These  spacetimes
also depend on two parameters,
and three exterior cases exist, depending on $\sigma$, the mass per unit length of the cylinder;
moreover,
the spacetimes contain CTC in the case of large $\sigma$ [2].

\section{Acknowledgements}
We are grateful to the E.P.S.R.C. for a Visiting Fellowship grant for
a visit to London by N.O.S. during which most of this work was completed.
We thank Prof. B.Carter, F.R.S. for correspondence about refs. [6] and [7],
and referees for advice about arrangement of the paper.

\section{References}
{[1]} van Stockum W J 1937 {\em Proc. Roy. Soc. Edin.} {\bf 57} 135\\
{[2]} Bonnor W B 1980 {\em J. Phys. A: Math. Gen.} {\bf13} 2121\\
{[3]} Santos N O 1993 {\em Class. Quantum Grav.} {\bf10} 2401\\
{[4]} Krasi\'{n}ski A 1975 {\em Acta Phys. Polon. B} {\bf 6} 223\\
{[5]} MacCallum M A H 1997 To be published in {\em Gen. Rel. Grav.}\\
{[6]} Carter B 1968 {\em Phys. Rev.} {\bf 174} 1559\\
{[7]} Carter B 1978 {\em Gen. Rel. Grav.} {\bf 9} 437\\
{[8]} Tipler F J 1974 {\em Phys. Rev. D} {\bf 9} 2203\\
{[9]} MacCallum M A H and Santos N O 1997 Submitted to {\em Class. Quantum Grav.}

\end{document}